\documentclass[twocolumn,showpacs,preprintnumbers,prl,aps,apssymb]{revtex4}

\usepackage{graphicx}
\usepackage{dcolumn}
\usepackage{bm}
\newcommand{\ber}{\begin{eqnarray}}
\newcommand{\eer}{\end{eqnarray}}
\newcommand{\bea}{\begin{equation}}
\newcommand{\eea}{\end{equation}}

\begin{document}


\title{A simple model for dynamic phase transitions in cell spreading}

\author{A. Bhattacharyay}
\affiliation{%
Department of Chemistry, University of Warwick, UK\\
}%

\date{\today}

\begin{abstract}
Cell spreading is investigated at various scales in order to understand motility of living cells which is essential for a range of physiological activities in higher organisms as well as in microbes. At a microscopic scale, it has been seen that actin polymerization at the leading edge of cell membrane primarily helps the cell to spread depending upon its extra-cellular environment which influences the polymerization process via some receptors on the cell membrane. There are some interesting experimental results at macroscopic scales (cell size) where people have observed various dynamic phases in terms of spreading rate of cell area adhering to the substrate. In the present paper we develop a very simple phenomenological model to capture these dynamic apparent phases of a spreading cell without going into the microscopic details of actin polymerization.         
\end{abstract}

\maketitle
\par
Living cells require to move to perform various activities. While microbes often move on a surface following a chemical gradient in search for food or for some other purpose, cells in multicellular organism also require to move for the purpose of immune-activity, healing, preferential positioning etc. When cells move on a solid substrate, they move by periodically spreading and contracting themselves. The microscopic phenomenon acting as a driving force for such motion has been understood to be actin polymerization \cite{Stossel,Pantaloni,Pollard,Ponti} against the lamellipodium at the leading edge of the spreading cell. Lamellipodium is basically an active gel enclosed by cell membrane. Inside this membrane, along the periphery of the spreading range, a network of actin filaments grow in outward direction and pushes the cell membrane forward while the actin filaments predominantly depolymerize at the inner side of the network to supply actin monomers for the outer spreading edge. Basically, due to depolymerization at the inner end and polymerization at the outer edge a local gradient of actin monomers form which keeps the process going and the mesh work moving in the outward direction. It has been observed that for the cell to spread properly, the binding of the actin filaments to the substrate on which it moves is quite important \cite{Linda}. This binding provides mechanical support to the actin network and thus helps the system avoid any breakup against mechanical restoring forces developed in the system as a result of spreading and also helps in the polymerization process by probably reducing positional fluctuations of the filaments. In this connection, the actin polymerization based motility of bacterium Listeria Monocytogene (LM) and bio-mimetic systems like actA coated polysterin beads are also worth mentioning \cite{Kuo,Lisa}. The LM, when invades a cell, captures the actin machinery of the cell and polymerizes an actin tail on one side of it. The continuous polymerization of the actins against the cell membrane of LM keeps it going on the opposite side. The depolymerization of the actin tail on the far end creates the local actin monomer gradient which supplies required actin monomers to the polymerizing end.
\par
With the advent of new powerful microscopy and imaging techniques people are now looking at this world of small scale biological activities at various levels. Particularly, in cell spreading experiments and theory people are trying to identify various universal features associated with the dynamics of spreading cells \cite{Doebereiner,Chamaraux,Doeb2}. Although, the process of cell spreading is a complex and active phenomenon involving complicated bio-mechanical pathways, efforts are on to look at the problem on the basis of simple physical principles without involving all the microscopic details. In \cite{Chamaraux}, it has been shown that the normalized contact area of a spreading cell to its substrate $<A(t)>/<A(t)_{t\rightarrow\infty}>$ is a universal function of time with a characteristic exponent which depend on cell type. The exponent $\alpha$ being a function of the cell type reflects differences in the physiology of cell types or probably differences in the environmental conditions upon which depends the spreading process, whereas, the same functional form reflects the basic underlying similarity of all the processes. In \cite{Chamaraux}, a model has been put forward to calculate the exponents from considerations of actin polymerization and depolymerization rates based on curvature of cell membrane at the leading edge, elastic properties of the substrate etc. Where, two distinct dynamical phases characterize the spreading of cells in ref. \cite{Chamaraux}, in ref. \cite{Doebereiner} three dynamical phases have been identified with the spreading of Mouse Embryonic Fibroblasts (MEF) on a fibronectin coated substrate. In the present paper we would propose a phenomenological model being based on gross experimental findings without going into any microscopic details in order to account for the existence of dynamical phases in a spreading cell. In the next section we will explain in some details the experiments in ref.\cite{Doebereiner} and the results. Following that we will propose our model and then will present the analysis of that model and results. Finally, we will conclude indicating the fact that the large scale phases of cell spreading probably do not crucially depend on all the microscopic varieties of the intricate bio-mechanical pathways rather are a manifestation of some gross mean field effects of all those molecular level intricacies.  

\par
In ref.\cite{Doebereiner}, various dynamic phases of a spreading cell on a suitable substrate have been investigated and a sequence of transitions between successive dynamic phases have been identified during the process of spreading. In this experiment MEF cells have been allowed to move on a glass surface coated with fibronetin. Fibronectin is an extra-cellular matrix protein which interacts with the cytoskeleton (an actin framework inside the cell) of the cell via the integrin receptors on the cell membrane. As has been mentioned above, this binding of the cytoskeleton to the extracellular matrix is very important for spreading of the cell by actin polymerization. A Total Internal Reflection Fluorescence microscopy and Differential Interference Contrast microscopy of the spreading cells revealed three distinct dynamic phases. In the initial phase the growth of the contact area of the cell is slow and has been seen to be characterize by a small growth exponent $a_1 = 0.4\pm 0.2$. It has been proposed that during this initial phase the cell basically tests the suitability of the surface to adhere on and once this testing time is gone the next phase of rapid growth of the area follows. This second phase has been characterized by a growth exponent $a_2 = 1.6\pm 0.9$. In the third phase the cell boundary shows periodic local contractions and the area of adherence to the substrate starts to oscillate and the mean area of contact increases very slowly untill it reaches the maximum limit. The growth exponent in this contractile expansion stage is $a_3=0.3\pm 0.2$.
\par
An important observation in \cite{Doebereiner} is that the cells taken in the experiment could be divided into two classes depending upon their growth rates in the middle fastest growing phases. Let the area of the cell at the point when the second dynamic phase starts from the first basal activity phase be $A_1$ and that at the point where the contractile phase takes it up from the rapid growth middle phase be $A_2$. All the cells in the experiment were found to belong to two classes depending upon the ratio $A_2/A_1<5$ or $A_2/A_1>5$. In the first class, the exponent in the middle phase was $a_2 = 0.9\pm 0.2$ whereas for the second class ($A_2/A_1>5$) it was $a_2 = 1.6\pm 0.2$. A large error bar appears in the measure of $a_2$ as mentioned in the previous paragraph is due to the fact that its a mean of $a_2$ in these two distinct classes. So, the experiment suggests that the bigger the maximum area of spread the faster is cell's growth and this is an important observation in order to write a phenomenological model for such systems. 
\par
Let us consider the model in the form
\begin{eqnarray}
\frac{\partial A}{\partial t}&=& \frac{1}{A}+pB-q\\\nonumber
\frac{\partial B}{\partial t}&=& r-A
\end{eqnarray}
where $A$ is the area in contact with the substrate and $B$ can be anything like polymerization rate at the leading edge of the cell membrane or any other form of active process that drives cell spreading and its growth rate depends upon availability of resources within the cell. Considering the cell as a closed system, the stock of above mentioned resources should get gradually reduced as a result of the spreading of the cell and that is why we have taken the growth rate of this encouragement factor to reduce with spread in the cell. For example, consider the $B$ to be polymerization rate at the leading edge. Since the growth of area will make the periphery of the cell grow this will in turn require supply of more actin monomers for further polymerization and growth. This supply should get eventually reduced since the cell is a closed system reducing the growth of polymerization rate as the cell spreads. The constant $r$ represents the limiting contact area of the spreaded cell for which the encouragement for spreading or the polymerization rate is just enough to maintain the dynamic equilibrium with the other degrading factors at that state. The growth rate of the area $A$ should depend on how big the $r$ is because a larger $r$ means larger initial rate of growth for the polymerization rate $B$ and that should make the contact area grow faster presumably in accordance with the experimentally observed facts. The first term on r.h.s of the equation for growth of the area $A$ is the one which stands for the spreading of the cell due to the pressure at the surface of contact with the substrate and other passive factors which get exhausted as the area of contact grows. For example, pressure to be proportional to the height $H$ and the volume of the cell being constant $H \sim 1/A$. The second term in the same equation is the one which represents the active process of cell spreading where $p$ is a constant. As has been mentioned in the ref.\cite{Doebereiner}, the cell initially takes some time to interact to the substrate in order to assess its suitability to be spreaded on; the constant $p$ in our  model has to be set very small in order to have the active spreading coming into effect when $B$ has grown by a good amount and up to that time the growth will be dominated by the other terms. The last term, a constant $q$ stands for all other things that constantly prevent spreading of the cell. 
\par
The fixed point of Eq.1 is given by $A_0=r$ and $B_0=(q-1/r)/p$. This fixed point actually corresponds to the final dynamic equilibrium state of the spreaded cell which has actually been spread to the limit where rate of polymerization  is equal to the rate of degradation of the actin filaments due to restoring forces. This can be easily understood if we do a linear stability analysis about this fixed point. Perturbing the system as $A=A_0+a$ and $B=B_0+b$ we have
\begin{eqnarray}
\frac{\partial a}{\partial t}&=& -\frac{a}{r^2}+pb\\\nonumber
\frac{\partial b}{\partial t}&=& -a
\end{eqnarray}  
The Growth rate of the perturbation is given by $\lambda = -1/r \pm \sqrt{1/r^2-4p}$, where the phase trajectory should spiral down to the fixed point ($A_0$,$B_0$) so long as $p>1/4r^2$. Such a phase portrait is shown in Fig.1 for $p=0.05$, $r=10$.
\par
The ideal initial values for our model will be $A$ very small and $B=0$. The area of the cell in contact with the substrate is taken to be small when it is just placed on the substrate. One can also consider that there is a threshold initial spreading $A_0$ say beyond which the spreading of the cell is effectively considered in the first term of Eq.1 and at the initial moment this excess area($A-A_0$) is very small and positive. Otherwise, one can also consider that the variable $A$ is the surface area in axcess to the initial area of spreading $A_0$ when the cell is just placed on the substrate and we actually have written $1/A$ instead of $1/(A_0+A)$ because the dynamics will basically remain the same in the region of our interest. Starting from such an initial condition the trajectories spiral down to the fixed point when $p>1/4r^2$ and we are interested in looking at the $A$ vs $t$ ($t$ is time) plot on log-log scales. Initially, when $B$ is very small and for very small value of the constant $p$ the growth rate of the area $A$ will effectively be given as
\begin{equation}
\frac{\partial A}{\partial t}= \frac{1}{A}-q
\end{equation}
which can be solved to get
\begin{equation}
A+\frac{1}{q}log(1-qA)=-qt
\end{equation}
This equation clearly shows that $A$ grows as $t^{1/2}$ for small enough $A$ and $q$ in the absence of contribution from active processes in spreading. This exponent $1/2$ corresponds quite well with the experimentally given one in ref.\cite{Doebereiner} $a_1=0.4\pm 0.2$. Such a spreading has been shown in Fig.2(a) for $r=10$, $p=0$ and $q=0.01$ . Now, keeping $p=0.01$, a small number we set the parameter $q$ at values 0.1 and 1.0 to plot the same in Fig.2(b) and (c). There are three distinct states of spreading as appear in these figures. The initial phase of a very small growth rate. In this phase the active part of our model i.e. ($pB$) has hardly any influence on the spreading process. Next comes a rapid growth phase where the system grows quite rapidly to a larger surface of contact followed by an oscillatory spreading phase. In the oscillatory spreading phase, the contact area not only oscillates but there is a small increase in the mean area of contact with time. This oscillatory growth or contractile growth process is better understood in the Fig.2(d) which has $p=0.05$ and $q=1.0$.
\par
The better manifestation of the contractile growth with a larger $p$ indicates that the contractile growth phenomenon is a characteristic of the competition between the active and other restoring passive processes in the system. In actual experiments, people have seen periodic breaking down of the actin mesh work at places along the circumference of the growing surface \cite{Ponti}. Such a local breakdown can happen as a result of development of mechanical stress as the cell spreads and supposedly some myosin density dependent generic contractile instability \cite{Kruse1,Kruse2}. To see if a bigger final area of contact corresponds to a larger growth exponent in the middle phase, we have plotted the same graph with $r=5,30$ for $p=.01$ and $q=1.0$ in Fig.3. The growth rate is clearly seen to increase with $r$ and our simple model qualitatively captures the experimental observations without going into microscopic details. The present analysis also indicates the fact that this increase in the exponent with the maximum area of contact should happen continuously rather than having classes of cells characterized by discrete exponents as is apparent from the experiment. This would be interesting to be further probed by experiments.
\par   
To conclude, we would like to mention that, the dynamic phases shown in cell spreading can easily be understood on the basis of dynamics of some macroscopic quantities and notion of conservations of ingredients in a closed system. The detailed understanding of the microscopic scale activities and their relations with the macroscopic parameters taken in the model are always important to realize the interplay between the small and large scale effects. One can always try to get the parameters used namely $p$, $q$ and $r$ in this minimal model from the microscopic level details of the  dynamics of the system or even get other new terms to be added to the minimal model in order to make it more appropriate, nevertheless, the information we get by qualitatively representing the experimental results with the simple phenomenological model is that the dynamics at the three different phases are really not that different. Its basically spiraling journey to a stable fixed point staring from a far falling initial state. The initial basal activity phase is definitely very much different from the other two in the sense that the active part of the dynamics is not appreciably present in that phase, but the middle phase of steep growth is basically the first half of the first period of oscillatory expansion. In our interpretation of the model, the high growth rate of the polymerization rate at the beginning when the contact area of the  cell with its substrate is small, is the cause of having this middle part as a separate phase on the log-log plot of the area against time. For the same dynamics, if the maximum area attainable by the spreading cell increases, it not only increase the growth rate of the system in this middle phase but would also reveal this middle phase to be a part of the integral contractile phase as is evident from the fig.3. This prediction can also be checked experimentally to understand the nature of these dynamic phases and assess the role of large scale (conserved) quantities on the controlling of cell spreading. The phenomenon of cell spreading is definitely not isotropic as has been considered in the present model. The contractile phase actually shows periodic local contraction along the circumference of the cell and lateral waves of some universal nature have been observed to appear at the circumference of the cell at this phase \cite{Doeb2}. Our simple isotropic model is really not in conflict with having local periodic contractions rather the universal spatio-temporal pattern of the lateral waves at the circumference of various cells indicates their common macroscopic origin. Thus, we conclude that the classification of cells in phenotypes depending upon their macroscopic behaviors during spreading over a suitable substrate crucially depends on the relative abundance of relevant ingredients that takes part in the microscopic process of spreading, or the gross difference in the elastic nature of the membranes or binding to substrate for different cell types rather on having subtle difference in the complex signaling pathways in different cells. 

\begin{figure}
\begin{center}
\includegraphics[scale=0.34,angle=-90]{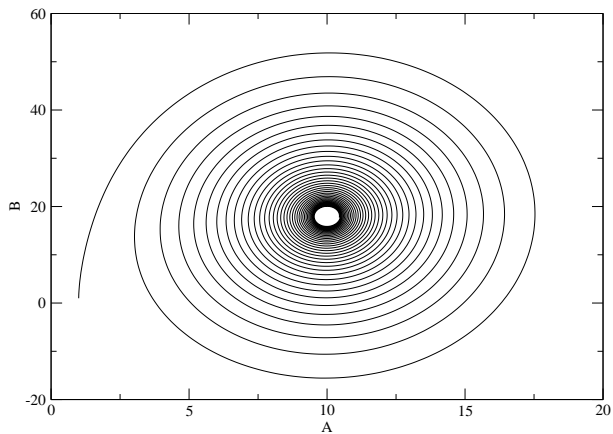}
\caption{Phase portrait showing the spiralling down to the fixed point ($A_0$,$B_0$).}
\label{Fig.1}
\end{center}       
\end{figure}
            
\begin{figure}
\begin{center}
\includegraphics[scale=0.34,angle=-90]{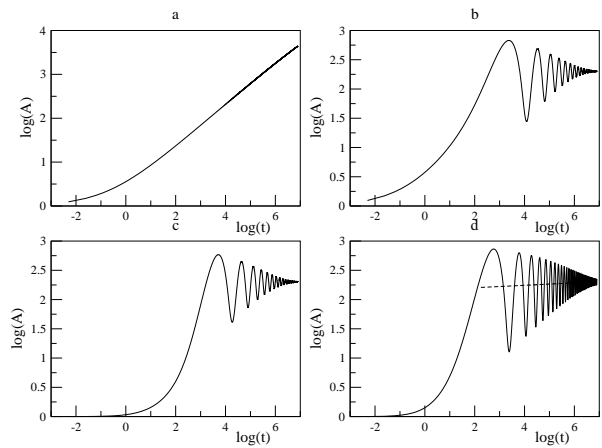}
\caption{Apparent dynamic phases shown by our model on a plot of log of area of contact $A$ against the logarithm of time $t$ while spreading from an initial state given by $A_{ini}=0.1$, $B_{ini}=0$.}
\label{Fig.2}
\end{center}       
\end{figure}    

\begin{figure}
\begin{center}
\includegraphics[scale=0.34,angle=-90]{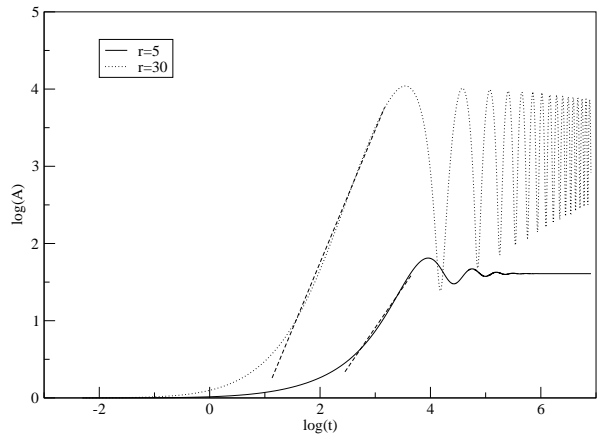}
\caption{Demonstration of the rapidity of spreading with respect to increase in the limiting spread $r$ of the cell. For $r=30$ the expontnt in the quickest spreading phase $a_2=1.72$ whereas $a_2=1.04$ when $r=5$.}
\label{Fig.2}
\end{center}       
\end{figure}

\end{document}